\documentclass[twocolumn,showpacs,preprintnumbers,amsmath,amssymb,showkeys]{revtex4}

\usepackage{graphicx}% Include figure files
\usepackage{dcolumn}% Align table columns on decimal point
\usepackage{bm}% bold math

\DeclareMathOperator{\tr}{tr}
\DeclareMathOperator{\Tr}{Tr}

\begin{document}

\title{Decoherence of Cooper pairs and subgap magnetoconductance of superconducting hybrids}
\author{Andrew G. Semenov$^{1}$
and Andrei D. Zaikin$^{2,1}$
}
\affiliation{$^1$I.E.Tamm Department of Theoretical Physics, P.N.Lebedev
Physics Institute, 119991 Moscow, Russia\\
$^2$Institute of Nanotechnology, Karlsruhe Institute of Technology (KIT), 76021 Karlsruhe, Germany
}

\begin{abstract}
We demonstrate that electron-electron interactions fundamentally restrict
the penetration length of superconducting correlations into a diffusive normal metal (N) attached to a superconductor (S). We evaluate the subgap magnetoconductance $G$ of SN hybrids in the presence of electron-electron
interactions and demonstrate that the effect of the magnetic field on $G$ is twofold: It includes ($i$) additional temperature independent dephasing of Cooper pairs and ($ii$) Zeeman splitting between the states with opposite spins. The dephasing length of Cooper pairs can be directly extracted
from measurements of the subgap magnetoconductance in SN systems at low temperatures.

\end{abstract}


\maketitle

\section{Introduction}

Cooper pairs may penetrate into a normal metal (N) from a superconductor (S) attached to this metal.
As a result, the latter can also sustain a non-vanishing supercurrent and demonstrate a number of other
properties inherent to a superconductor \cite{dG,Tink}. This superconducting proximity
effect may be interpreted in terms of the well known phenomenon called ``Andreev reflection'' \cite{And}:
Cooper pairs reaching the NS interface get converted into subgap quasiparticles which further
diffuse into the normal metal maintaining the information about a macroscopic
phase of the superconducting condensate. At non-zero temperatures this macroscopic
quantum coherence of single electrons in the N-metal can, however, be destroyed by thermal
fluctuations at a typical length of order $L_T \sim \sqrt{D/T}$, where
$D$ is the electron diffusion coefficient.  It follows immediately that
at sufficiently low temperature the whole normal
metal can become essentially superconducting and such
phenomena as, e.g., Meissner effect in NS strucures \cite{Z82}, Josephson effect in
SNS junctions \cite{Kulik,Ishii} as well as
subgap electron transport across NS interfaces \cite{BTK} can be observed.
Interesting features of the latter phenomenon will be investigated in our present work.

At sufficiently low values of the NS interface
transmission its subgap (Andreev) conductance
$G$ is proportional to the second order in the barrier transmission \cite{BTK}
and, hence, typically it remains rather small. On the other hand,
in the low energy limit $G$ can be strongly enhanced
due to non-trivial interplay between disorder and quantum
interference of electrons in the normal metal \cite{VZK,HN,Ben,Zai}.
In the limit of low voltages and temperatures this effect yields
the so-called zero-bias anomaly (ZBA) $G \propto 1/\sqrt{V}$
and $G \propto 1/\sqrt{T}$.

It is important to note that all these results remain applicable only provided electron-electron interactions in the N-metal can be totally disregarded. Though in some cases this assumption is indeed justified, in general Coulomb effects
can play a significant role and need to be properly accounted for.
Various aspects of subgap electron
transport across NS interfaces in the presence of such effects were studied in a number of papers \cite{Zai,HHK,ZGalakt,ZGalakt2,SZK12}.

For instance, it was demonstrated \cite{SZK12}
that electron-electron interactions
yield {\it dephasing of Cooper pairs}
in the normal metal which, in turn, can significantly influence the subgap conductance of NS systems. In simple terms this phenomenon can be interpreted as an effect of
fluctuating electromagnetic field produced by fluctuating electrons in a disordered
normal metal. It turns out that such fluctuating field destroys macroscopic coherence of electrons penetrating from a
superconductor at a certain characteristic length $L_{\varphi}$, thereby imposing fundamental limitations on the proximity effect
in NS hybrids at low temperatures $T \lesssim D/L_\varphi^2$. At such
temperatures the penetration depth
of superconducting correlations into the N-metal is not anymore determined
by the thermal length $L_T$, but is limited by the dephasing length $L_\varphi$ which -- unlike $L_T$ -- does not diverge at $T \to 0$.

It is also worthwhile to point out that the dephasing length $L_\varphi$
obtained  for NS systems \cite{SZK12} up to a numerical prefactor coincides with zero temperature decoherence length derived within a completely different theoretical
framework \cite{GZ1,GZ3,GZ4,GZS,GZ5} for the
weak localization (WL) correction to the conductivity of a disordered normal metal. This
agreement emphasizes fundamental and universal
nature of low temperature dephasing by electron-electron interactions in different types of disordered conductors, including
NS hybrids analyzed here. At the same time, it was demonstrated \cite{SZK12} that dephasing of Cooper pairs by
electron-electron interactions is in several important aspects different from
that for single electrons in N-metals. These aspects will also be illustrated below.

In this work we will theoretically analyze a combined effect of electron-electron interactions and of an external magnetic field on subgap electron transport in diffusive SN structures. We will
demonstrate that low temperature transport experiments with SN hybrids allow to directly probe the dephasing length of Cooper pairs in such systems.

\begin{figure}[h]
\includegraphics[width=0.7\columnwidth]{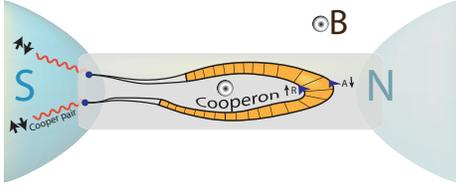}
\caption{Hybrid SN structure in an external magnetic field $B$ and the diagram describing conversion of a Cooper pair into a pair of
electrons propagating inside the normal metal.}
\label{fig1}
\end{figure}

\section{The model and formalism}

In what follows we will analyze a hybrid SN structure which
consists of a normal metallic wire with cross-section
$a^2$ and length $L \gg a$  attached to bulk superconducting and normal
electrodes, see Fig. 1.
We will assume that the normal wire and the superconducting electrode are
connected via a tunnel barrier of the same cross-section $a^2$ and resistance
$R_{\rm I}$ strongly exceeding the wire resistance $R_{\rm I}\gg R=L/(\sigma a^2)$, where
$\sigma=2e^2\nu D$ is the wire Drude conductivity, $e$ is the electron charge
and $\nu$ is the density of states per spin direction. We
will also assume that a comparatively weak uniform magnetic field $B$ is applied to the system.

For our analysis we will employ the Keldysh version of the nonlinear $\sigma$-model \cite{KA,KA1} extended to describe SN structures
\cite{ZGalakt,SZK12}. The Keldysh effective action for such systems
defined on the time contour with forward (F) and backward (B) parts is expressed as a sum of two terms
$S=S_w+S_{\rm I}$ which account respectively for diffusive motion of electrons
in the wire,
\begin{equation}
S_w[\check Q,{\bf A},\Phi]=\frac{i\pi\nu}{4}\Tr[D(\check\partial \check Q)^2-\check\Xi(4\partial_t+2i\omega_Z)\check Q+4i\check\Phi\check Q],
\label{swire}
\end{equation}
and electron tunneling across the junction \cite{KL},
\begin{equation}
S_\Gamma[\check Q,{\bf A},\Phi]=-\frac{i\pi}{4e^2
  R_{\rm I}a^2}\Tr_{\rm I}[\check Q_{\rm sc}\check Q],
\label{sI}
\end{equation}
where $\check Q_{\rm sc}$ and $\check Q$ are defined on respectively  superconducting and normal
sides of the insulating barrier, the quantity $\omega_Z=g\mu_BB$ (with $g\mu_B$ being the product of a $g$-factor and the Bohr magneton) accounts for Zeeman splitting,  and "$\Tr$"\  indicates the
trace over the matrix indices and the integration over both time
and coordinate variables.  The covariant derivative is defined as
\begin{equation}
\check\partial \check Q=\partial_{{\bf r}}\check Q-i[\check\Xi\check {\bf
A},\check Q], \quad \check\Xi=\left(\begin{array}{cc} \hat \sigma_z &  0 \\
0 & \hat\sigma_z \end{array}\right),
\end{equation}
where $[x,y]$ is the commutator and $\hat\sigma_{x,y,z}$ denotes the set of Pauli matrices.
Both parts of the action (\ref{swire}) and
(\ref{sI}) contain the $4\times4$ dynamical matrix field $\check Q$ satisfying the standard
normalization condition $\check Q^2 =\check 1\delta(t-t')$
as well as  the scalar and vector potentials
$\Phi({\bf r},t)$ and ${\bf A}({\bf r},t)$ which account for the effect of electron-electron interactions as well as the effect of an external magnetic field $H$. These potentials are defined on the Keldysh
contour and, hence, one can introduce the variables
$\Phi^{\pm}=\frac{1}{\sqrt{2}}(\Phi^F\pm\Phi^B)$ and  ${\bf
  A}^{\pm}=\frac{1}{\sqrt{2}}({\bf A}^F\pm{\bf A}^B)$ and define the matrices
\begin{equation}
\check\Phi =\left(\begin{array}{cc} \Phi^+\hat1 & \Phi^-\hat1  \\
 \Phi^-\hat1 & \Phi^+ \hat1
 \end{array}\right), \quad\check{\bf A}=\left(\begin{array}{cc} {\bf A}^+\hat1 & {\bf A}^-\hat1  \\
 {\bf A}^-\hat1 & {\bf A}^+ \hat1
 \end{array}\right).
\end{equation}

As we are merely interested in evaluation of the subgap (Andreev) conductance
it suffices to restrict our consideration to
energies well below the superconducting gap. In this limit we set
\begin{equation}
\check Q_{\rm sc}(t,t')=\left(\begin{array}{cc}\hat \sigma_y &
0\\0&\hat\sigma_y\end{array}\right)\delta(t-t').
\end{equation}
In order to proceed we will employ the so-called $\mathcal K$-gauge trick \cite{KA,KA1} and perform the gauge transformation
$\check Q({\bf r},t,t')\to e^{i\check\Xi \check{\mathcal K}({\bf r},t)}\check Q({\bf r},t,t') e^{-i\check\Xi \check{\mathcal K}({\bf r},t')}$
which eliminates the linear terms in both electromagnetic potentials and deviations from the normal metal saddle point
\begin{gather}
\check Q_N=\check {\mathcal U}\circ \left(\begin{array}{cc}
\hat\sigma_z&0\\0&-\hat\sigma_z
\end{array}\right)\check {\mathcal U},\\
%\end{equation}
%\begin{equation}
\check{\mathcal U}(t-t')=\left(\begin{array}{cc}\delta(t-t'-0)\hat 1 &
-\frac{iT}{\sinh(\pi T(t-t'))}\hat 1 \\ 0 &
-\delta(t-t'+0)\hat1\end{array}\right).
\end{gather}
This goal is achieved if one chooses the $\mathcal K$-field to obey the following equations
%\begin{multline}
\begin{eqnarray}
 \Phi_{\mathcal K}^+({\bf r},t)&=&D\partial_{\bf r}{\bf A}_{\mathcal K}^+({\bf r},t) \\
&&-2iDT\int dt'\coth(\pi T(t-t'))\partial_{\bf r}{\bf A}_{\mathcal K}^-({\bf r},t'),\nonumber\\
%\end{multline}
%\begin{equation}
 \Phi_{\mathcal K}^-({\bf r},t)&=&-D\partial_{\bf r}{\bf A}_{\mathcal K}^-({\bf r},t)
\end{eqnarray}
with $\Phi_{\mathcal K}({\bf r},t)=\Phi({\bf r},t)-\partial_t\mathcal K({\bf r},t)$ and ${\bf A}_{\mathcal K}({\bf r},t)={\bf A}({\bf
r},t)-\partial_{\bf r}\mathcal K({\bf r},t)$. As a result of this transformation the total action retains its initial form provided one substitutes
$  \check Q_{\rm sc}(t,t')\to e^{-i\check\Xi \check{\mathcal K}({\bf r},t)}\check Q_{\rm sc}(t,t')e^{i\check\Xi \check{\mathcal K}({\bf r},t')}$, $\Phi\to\Phi_{\mathcal K}$ and ${\bf A}\to{\bf A}_{\mathcal K}$.

Treating the tunneling term (\ref{sI}) perturbatively and performing the integration over the $\check Q$-field, analogously to \cite{ZGalakt,SZK12} we
recover the so-called Andreev contribution to the effective action
\begin{equation}
S_A=-\frac{i}{32}\left(\frac{\pi}{e^2R_{\rm I} a^2}\right)^2\langle \Tr_{\rm I}[\check Q_{\rm sc}\check Q]\Tr_{\rm I}[\check Q_{\rm sc}\check Q]\rangle_Q,
\label{sa}
\end{equation}
which accounts for all processes of Andreev reflection at the SN interface. The dependence of $S_A$ on the electromagnetic potentials is contained both in $\check Q_{\rm sc}$ and in the average of the $\check Q$-fields. In order to evaluate this average it is necessary to parametrize the deviations of the $\check Q$-matrix from the metallic saddle point $\check Q_N$.  Here we will make use of the parametrization \cite{KA,KA1}
%\begin{equation}
$
 \check Q\approx\check Q_N
 +i\check Q_N\circ \check{\mathcal U}\circ\check { W}\circ \check{\mathcal U}-\frac12\check Q_0\circ\check {\mathcal U}\circ\check { W}\circ\check { W}\circ\check {\mathcal U}+O(\check W^4),
$
%\end{equation}
where
\begin{equation}
\check W =\left(\begin{array}{cccc} 0 &c^{(1)} ({\bf r},t,t') & d^{(1)} ({\bf r},t,t') & 0 \\
\bar c^{(1)} ({\bf r},t',t)& 0& 0 & d^{(2)} ({\bf r},t,t') \\
\bar d ^{(1)}({\bf r},t',t)& 0& 0 & c^{(2)} ({\bf r},t,t') \\
0&\bar d^{(2)} ({\bf r},t',t)&  \bar c^{(2)}({\bf r},t',t) & 0
\end{array}\right)\nonumber
\end{equation}
and $d^{(j)}$, $c^{(j)}$ are respectively the diffuson and the Cooperon fields.  Inserting this expansion into $S_w$ and collecting all terms up to the second order in $\check W$ one arrives at the effective action for diffusons and Cooperons interacting with the electromagnetic field. Higher order terms  of the above expansion generate interactions between diffusons and Cooperons and will be ignored further below. We also note that turning from the integration over the $\check Q$-variable to that over $\check W$-variable one introduces the Jacobian which -- in the above parametrization -- does not influence the second order terms. We will get back to this discussion in the next section.

In order to evaluate the Andreev action it is sufficient to keep only the first order terms in $\check W$ under each trace which only depend on the Cooperon fields. Introducing the notations
\begin{equation}
\hat c^{(j)}_{\varepsilon,\varepsilon'}({\bf r})=\int dtdt' e^{i\varepsilon t-i\varepsilon't'}
\left(\begin{array}{cc} 0 &c^{(j)} ({\bf r},t,t')  \\
\bar c^{(j)} ({\bf r},t',t)& 0
\end{array}\right),
\end{equation}
\begin{equation}
\hat\alpha_{\omega}({\bf r})=\int dt e^{i\omega t}\left(\begin{array}{cc} 0 &-ie^{-2i\mathcal K^+} \\
ie^{2i\mathcal K^+}& 0
\end{array}\right)\cos(2\mathcal K^-),
\end{equation}
\begin{equation}
\hat\beta_{\omega}({\bf r})=\int dt e^{i\omega t}\left(\begin{array}{cc} 0 &-e^{-2i\mathcal K^+} \\
-e^{2i\mathcal K^+}& 0
\end{array}\right)\sin(2\mathcal K^-),
\end{equation}
we rewrite the trace in the form
\begin{multline}
\Tr_{\rm I}[\check Q_{\rm sc},\check Q]\approx i\int\limits_{\rm I}d{\bf r}\int\frac{d\varepsilon d\omega}{(2\pi)^2}\tr[(\hat \alpha_\omega({\bf r})+F_{\varepsilon+\omega}\hat\beta_\omega({\bf r}))
\\\times\hat\sigma_z(\hat c^{(1)}_{\varepsilon,\varepsilon+\omega}({\bf r})-\hat c^{(2)}_{-\varepsilon-\omega,-\varepsilon}({\bf r}))],
\label{trc}
\end{multline}
where we defined $F_\varepsilon=\tanh\frac{\varepsilon}{2T}$. One observes that this expression depends only on the combination
\begin{equation}
\hat c^{as}_{\varepsilon,\varepsilon'}({\bf r})=\frac{\hat c^{(1)}_{\varepsilon,\varepsilon'}({\bf r})-\hat c^{(2)}_{-\varepsilon',-\varepsilon}({\bf r})}{\sqrt{2}},
\label{asymm}
\end{equation}
which has a clear physical meaning. One can show that the Cooperon can be represented as a sum of the impurity ladder diagrams involving retarded ($G^R$) and advanced ($G^A$) Green functions. As Cooper pairs are spin-singlets,  the spin structure of the corresponding Cooperon relevant for the proximity induced superconductivity is either $\uparrow\downarrow$ or $\downarrow\uparrow$. One can check that $\hat c^{(1)}$($\hat c^{(2)}$) Cooperon field corresponds to the  $\downarrow\uparrow(\uparrow\downarrow)$ spin configuration. It follows from Eq. (\ref{trc}) that only antisymmetric combination enters into all the expressions derived in the leading order. Note that the Cooperon analyzed here differs from that of the Cooperon encountered,  e.g., in the WL problem, as the latter is described either by $\uparrow\uparrow$ or by $\downarrow\downarrow$ spin configuration. In the next section we will demonstrate that -- depending on its spin structure -- the Cooperon behavior can differ substantially already at the level of the first order perturbation theory in the electron-electron interactions.

\section{Perturbation theory for the Cooperon}

We proceed by expanding the action $S_w$ up to the second order in both the diffuson and the Cooperon fields. Then we get
\begin{equation}
S_w=S_w^{(0,2)}+S_w^{(1,2)}+S_w^{(2,1)}+S_w^{(2,2)},
\end{equation}
where the terms $S^{(i,j)}$ contain $i$-th power of the electromagnetic
potentials and $j$-th power of $\check W$. It is straightforward to verify that the term $S^{(2,1)}$ depends only on the diffuson fields
and, hence, is irrelevant for the problem under consideration. The remaining terms read
\begin{multline}
S_w^{(0,2)}=\frac{i\pi\nu}{4}\sum_{j=1,2}\Tr[D(\partial_{\bf r}\hat c_{\varepsilon,\varepsilon'}^{(j)})(\partial_{\bf r}\hat c_{\varepsilon',\varepsilon}^{(j)})
\\+(-1)^j(2i\varepsilon-i\omega_Z)\hat c_{\varepsilon,\varepsilon'}^{(j)}\hat c_{\varepsilon',\varepsilon}^{(j)}],
\end{multline}
\begin{widetext}
\begin{multline}
S_{w}^{(1,2)}=-\pi\nu D\Tr[\hat c^{(1)}_{\varepsilon,\varepsilon'+\omega} {\bf A}^{+}_{\mathcal K,\omega}\hat\tau_z(\partial_{\bf r}\hat c^{(1)}_{\varepsilon',\varepsilon})-(\partial_{\bf r}\hat c^{(2)}_{\varepsilon,\varepsilon'+\omega}) {\bf A}^{+}_{\mathcal K,\omega}\hat\tau_z\hat c^{(2)}_{\varepsilon',\varepsilon}
-(\partial_{\bf r}\hat
c^{(1)}_{\varepsilon,\varepsilon'+\omega})F_{\varepsilon'+\omega} {\bf A}^{-}_{\mathcal K,\omega}\hat\tau_z\hat c^{(1)}_{\varepsilon',\varepsilon}\\-\hat
c^{(2)}_{\varepsilon,\varepsilon'+\omega} {\bf A}^{-}_{\mathcal K,\omega}F_{\varepsilon'}\hat\tau_z(\partial_{\bf r}\hat
c^{(2)}_{\varepsilon',\varepsilon})-\hat
c^{(1)}_{\varepsilon,\varepsilon'+\omega}B_\omega (\partial_{\bf r}{\bf A}^{-}_{\mathcal K,\omega})\hat\tau_z\hat c^{(1)}_{\varepsilon',\varepsilon}+\hat
c^{(2)}_{\varepsilon,\varepsilon'+\omega}B_\omega (\partial_{\bf r}{\bf A}^{-}_{\mathcal K,\omega})\hat\tau_z\hat c^{(2)}_{\varepsilon',\varepsilon}]
\end{multline}
and
\begin{multline}
S_{w}^{(2,2)}=\frac{i\pi\nu D}{2}\Tr[({\bf A}^{+}_{\mathcal K,\omega_1}{\bf A}^{+}_{\mathcal K,\omega_2}+2{\bf A}^{+}_{\mathcal K,\omega_1}F_{\varepsilon+\omega_2}{\bf A}^{-}_{\mathcal K,\omega_2}-{\bf A}^{-}_{\mathcal K,\omega_1}{\bf A}^{-}_{\mathcal K,\omega_2}+2F_{\varepsilon+\omega_1+\omega_2}{\bf A}^{-}_{\mathcal K,\omega_1}F_{\varepsilon+\omega_2}{\bf A}^{-}_{\mathcal K,\omega_2})\hat c^{(1)}_{\varepsilon,\varepsilon'}\hat c^{(1)}_{\varepsilon',\varepsilon+\omega_1+\omega_2}\\
+({\bf A}^{+}_{\mathcal K,\omega_1}{\bf A}^{+}_{\mathcal K,\omega_2}-2{\bf A}^{-}_{\mathcal K,\omega_1}F_{\varepsilon+\omega_2}{\bf A}^{+}_{\mathcal K,\omega_2}-{\bf A}^{-}_{\mathcal K,\omega_1}{\bf A}^{-}_{\mathcal K,\omega_2}+2{\bf A}^{-}_{\mathcal K,\omega_1}F_{\varepsilon+\omega_2}{\bf A}^{-}_{\mathcal K,\omega_2}F_{\varepsilon})\hat c^{(2)}_{\varepsilon,\varepsilon'}\hat c^{(2)}_{\varepsilon',\varepsilon+\omega_1+\omega_2}\\+ ({\bf A}^{+}_{\mathcal K,\omega_1}+F_{\varepsilon+\omega_1}{\bf A}^{-}_{\mathcal K,\omega_1})\hat c^{(1)}_{\varepsilon,\varepsilon'+\omega_2}({\bf A}^{+}_{\mathcal K,\omega_2}+F_{\varepsilon'+\omega_2}{\bf A}^{-}_{\mathcal K,\omega_2})\hat c^{(1)}_{\varepsilon',\varepsilon+\omega_1}\\-2{\bf A}^{-}_{\mathcal K,\omega_1}\hat c^{(1)}_{\varepsilon,\varepsilon'+\omega_2}({\bf A}^{+}_{\mathcal K,\omega_2}+B_{\omega_2}{\bf A}^{-}_{\mathcal K,\omega_2})(F_{\varepsilon'+\omega_2}-F_{\varepsilon'})\hat c^{(2)}_{\varepsilon',\varepsilon+\omega_1}\\+({\bf A}^{+}_{\mathcal K,\omega_1}-{\bf A}^{-}_{\mathcal K,\omega_1}F_{\varepsilon})\hat c^{(2)}_{\varepsilon,\varepsilon'+\omega_2}({\bf A}^{+}_{\mathcal K,\omega_2}-{\bf A}^{-}_{\mathcal K,\omega_2}F_{\varepsilon'})\hat c^{(2)}_{\varepsilon',\varepsilon+\omega_1}].
\end{multline}
\end{widetext}

\begin{figure}[h]
\includegraphics[width=0.7\columnwidth]{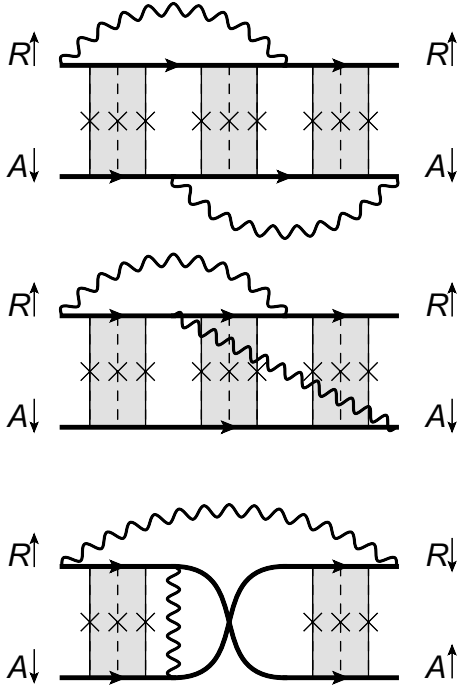}
\caption{Three different types of diagrams which contribute to the self-energy of the Cooperon. The dashed lines represent scattering on impurities and the wavy line accounts for electron-electron interactions.}
\label{fig2}
\end{figure}

Here we also defined $B_\omega=\coth\frac{\omega}{2T}$. Making use of these expressions one can proceed with the perturbation theory. Defining the Cooperons as
\begin{multline}
2\pi\delta(\omega-\omega') \mathcal C_{ij}({\bf r}-{\bf r}';\omega;\varepsilon,\varepsilon')=
\frac{\pi\nu}{2}\\\times\Big\langle \bar c^{(i)}_{\varepsilon+\frac{\omega}{2},\varepsilon-\frac{\omega}{2}}\left({\bf r}\right)
c^{(j)}_{\varepsilon'+\frac{\omega}{2},\varepsilon'-\frac{\omega}{2}}\left({\bf r}\right) \Big\rangle_{Q,\Phi},
\label{coop}
\end{multline}
where $c^{(j)}_{\varepsilon,\varepsilon'}({\bf r})=\int dt dt' e^{i\varepsilon t-i\varepsilon't'}c^{(j)}({\bf r},t,t')$, one can derive the corresponding
Dyson equation with the self-energy that consists of three different contributions. The first one, $S_d({\bf r}-{\bf r}';\omega;\varepsilon)$, turns out to be
proportional to $2\pi\delta(\varepsilon-\varepsilon')$. On the diagrammatic level it originates from the diagrams with retarded and advanced lines connected
only by the impurity lines. The second contribution $S_c({\bf r}-{\bf r}';\omega;\varepsilon,\varepsilon')$ is expressed as a sum of all irreducible diagrams
with non-crossed retarded and advanced lines connected by interactions.  Finally, the the third contribution
$S_{cr}({\bf r}-{\bf r}';\omega;\varepsilon,\varepsilon')$ corresponds to diagrams with crossed retarded and advanced lines.
Examples of these different types of diagrams are displayed in Fig. 2.

For simplicity let us consider an infinite system and take the limit $B \to 0$.
In this case the Dyson equation for the Cooperon can be written as
\begin{widetext}
\begin{equation}
\mathcal C_{ij}({\bf p};\omega;\varepsilon,\varepsilon')=2\pi\delta(\varepsilon-\varepsilon')\mathcal C_{ij}^{(0)}({\bf p};\varepsilon)+\frac{2}{\pi\nu}\sum_{k,l=1,2}\mathcal C_{ik}^{(0)}({\bf p};\varepsilon)\int\frac{d\varepsilon''}{2\pi}\mathcal S_{kl}({\bf p};\omega;\varepsilon,\varepsilon'')\mathcal C_{lj}({\bf p};\omega;\varepsilon'',\varepsilon'),
\end{equation}
where
\begin{equation}
\mathcal S_{ij}({\bf p};\omega;\varepsilon,\varepsilon')=\left(\begin{array}{cc}
2\pi\delta(\varepsilon-\varepsilon')S_d({\bf p};\omega;\varepsilon)+S_c({\bf p};\omega;\varepsilon,\varepsilon') & S_{cr}({\bf p};\omega;\varepsilon,-\varepsilon') \\
S_{cr}({\bf p};\omega;-\varepsilon,\varepsilon') &2\pi\delta(\varepsilon-\varepsilon')S_d({\bf p};\omega;-\varepsilon)+S_c({\bf p};\omega;-\varepsilon,-\varepsilon')
\end{array}\right),
\end{equation}
\end{widetext}
\begin{equation}
\mathcal C_{ij}^{(0)}({\bf p};\varepsilon) =\left(\begin{array}{cc}
\frac{1}{-2i\varepsilon+D{\bf p}^2} & 0 \\
0 &\frac{1}{2i\varepsilon+D{\bf p}^2}
\end{array}\right).
\end{equation}
Note that while deriving these equations we employed the symmetry of the term $S_{w}$.

The above equations can be diagonalized by expressing them in terms of the two fields $\hat c^{s}$ and $\hat c^{as}$, where a symmetric one is
\begin{equation}
\hat c^{s}_{\varepsilon,\varepsilon'}({\bf r})=\frac{\hat c^{(1)}_{\varepsilon,\varepsilon'}({\bf r})+\hat c^{(2)}_{-\varepsilon',-\varepsilon}({\bf r})}{\sqrt{2}},
\label{symm}
\end{equation}
whereas an antisymmetric field was already defined in Eq. (\ref{asymm}). With thin in mind and making use of Eq. (\ref{coop}) we introduce a symmetric and an
antisymmetric versions of the
Cooperon, respectively $\mathcal C_{s}$ and $\mathcal C_{as}$, which satisfy the following integral equations
\begin{multline}
(-2i\varepsilon+D{\bf p}^2)\mathcal C_{s}({\bf p};\omega;\varepsilon,\varepsilon')=2\pi\delta(\varepsilon-\varepsilon')
\\+\frac{2}{\pi\nu}\int\frac{d\varepsilon''}{2\pi}\mathcal S_{c}({\bf p};\omega;\varepsilon,\varepsilon'')\mathcal C_{s}({\bf p};\omega;\varepsilon'',\varepsilon')
\\+\frac{2}{\pi\nu}\int\frac{d\varepsilon''}{2\pi}\mathcal S_{cr}({\bf p};\omega;\varepsilon,\varepsilon'')\mathcal C_{s}({\bf p};\omega;\varepsilon'',\varepsilon')
\\+\frac{2}{\pi\nu}S_d({\bf p};\omega;\varepsilon)\mathcal C_{s}({\bf p};\omega;\varepsilon,\varepsilon'),
\end{multline}
\begin{multline}
(-2i\varepsilon+D{\bf p}^2)\mathcal C_{as}({\bf p};\omega;\varepsilon,\varepsilon')=2\pi\delta(\varepsilon-\varepsilon')
\\+\frac{2}{\pi\nu}\int\frac{d\varepsilon''}{2\pi}\mathcal S_{c}({\bf p};\omega;\varepsilon,\varepsilon'')\mathcal C_{as}({\bf p};\omega;\varepsilon'',\varepsilon')
\\-\frac{2}{\pi\nu}\int\frac{d\varepsilon''}{2\pi}\mathcal S_{cr}({\bf p};\omega;\varepsilon,\varepsilon'')\mathcal C_{as}({\bf p};\omega;\varepsilon'',\varepsilon')
\\+\frac{2}{\pi\nu}S_d({\bf p};\omega;\varepsilon)\mathcal C_{as}({\bf p};\omega;\varepsilon,\varepsilon')
\end{multline}
Note that an extra minus sign in front of one of the terms in the second equation arises because each crossing of retarded and advanced lines exchanges the spins.

It is straightforward to demonstrate that the symmetric version of the Cooperon $\mathcal C_{s}$ coincides with that encountered, e.g., within the WL problem.
At the same time, the antisymmetric Cooperon $\mathcal C_{as}$ responsible for the proximity effect studied here turns out to be different.
Hence, in general the results derived for the WL problem \cite{GZ1,GZ3,GZ4,GZS,GZ5} do not yet allow one to draw any definite conclusion about dephasing of
Cooper pairs by
electron-electron interactions. A striking difference between  $\mathcal C_{s}$ and $\mathcal C_{as}$ is observed already in the first order perturbation theory
in the interactions. While the first order diagrams for the Cooperon $\mathcal C_{s}$ (see Fig. 3) cancel each other exactly in the limit
$T \to 0$, $\omega \to 0$ and ${\bf p} \to 0$, the same diagrams for the Cooperon $\mathcal C_{as}$ do not cancel at all in this limit.

\begin{figure}[h]
\includegraphics[width=0.7\columnwidth]{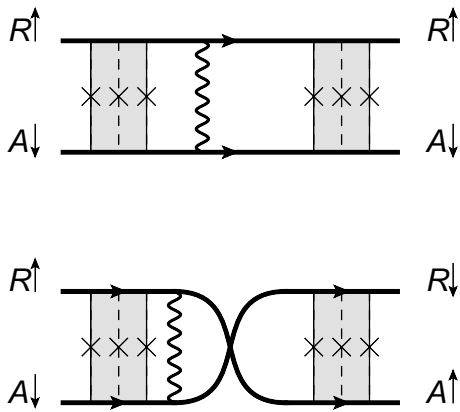}
\caption{The diagrams representing the first order interaction corrections to both symmetric and antisymmetric Cooperons $\mathcal C_{s}$ and $\mathcal C_{as}$.}
\label{fig3}
\end{figure}

The above observation implies that -- unlike in the WL problem -- non-vanishing zero temperature dephasing of Cooper pairs occurs already in the first
order in the electron-electron interactions. If one is tempted to estimate the magnitude of the dephasing effect encountered within the first order
perturbation theory, one could -- assuming purely exponential decay of correlations -- simply exponentiate the first order result and
arrive at the the equation
\begin{equation}
4D\int\frac{d\varepsilon d^d{\bf q}}{(2\pi)^{d+1}} \frac{\langle \mathcal K^+_\omega({\bf q})
\mathcal K^-_{-\omega}(-{\bf q}) \rangle {\bf q}^2\varepsilon\tau_\varphi}{|\varepsilon|(1-2i\varepsilon\tau_{\varphi}^{\rm pert})}\approx 1,
\end{equation}
which defines the perturbatively evaluated zero temperature dephasing time for Cooper pairs  $\tau_\varphi^{\rm pert}$.
In the universal limit of strong interactions this equation yields $\tau_\varphi^{\rm pert}\approx 8\nu^2Da^4$.

Note that this first order result  cannot be considered as a correct one simply because all higher order terms of the perturbation theory
need to be taken into account. As it was repeatedly explained elsewhere \cite{GZ3,GZ4,GZS,SZK12}, the problem in question is essentially non-perturbative
and, hence, it is an imperative to sum up the perturbation theory to {\it all} orders. This goal is accomplished within the semiclassical analysis in
the next section.

\section{Nonperturbative analysis}

The non-perturbative semiclassical approach in the Keldysh technique amounts to expanding the exact effective action in
the so-called quantum components of the fluctuating fields defined as a difference between the fields on forward and backward parts of Keldysh contour.
This approximation is effectively equivalent to dropping Coulomb blockade effects which are negligible
in the limit of large conductances of a normal wire $g_w=4\pi\nu Da^2/L \gg 1$ except at exponentially low voltages and temperatures.
E.g. one can show \cite{ZGalakt} that in this limit the terms  $\langle \mathcal K^+\mathcal K^-\rangle$
give negligible contribution to Andreev conductance as compared to that of the terms $\langle \mathcal K^+\mathcal K^+\rangle$.
Thus, in the main approximation one can simply drop the fields $\Phi^-$, $\mathcal K^-$, and ${\bf A}^-$ and
represent the Cooperons (before averaging over fluctuating electromagnetic fields)
\begin{multline}
\delta(T-T') \mathcal C_i({\bf r},{\bf r}';T;\tau,\tau')=\frac{\pi\nu}{2}\\\times\Big\langle \bar c^{(i)}\left({\bf r},T-\frac{\tau}{2},T+\frac{\tau}{2}\right) c^{(i)}\left({\bf r},T'-\frac{\tau'}{2},T'+\frac{\tau'}{2}\right) \Big\rangle_{Q}
\label{coop1}
\end{multline}
as a solution of the diffusion-like equation
\begin{widetext}
\begin{multline}
\left( 2\partial_\tau-i\Phi_{\mathcal K}^+({\bf r},T-\tau/2)+i\Phi_{\mathcal K}^+({\bf r},T+\tau/2)-D(\partial_{\bf r}+i{\bf A}^+_{\mathcal K}({\bf r},T-\tau/2)+i{\bf A}^+_{\mathcal K}({\bf r},T+\tau/2))^2+i\omega_Z \right)\mathcal C_1({\bf r},{\bf r}';T;\tau,\tau')\\=\delta({\bf r-r'})\delta(\tau-\tau')
\label{coop2}
\end{multline}
Resolving Eq. (\ref{coop2}) in presence of an external weak magnetic field $B$, for the antisymmetric Cooperon we obtain
\begin{multline}
\mathcal C_{as}({\bf r},{\bf r}';T;\tau,\tau')=
\frac{1}{2}\theta(\tau-\tau')\cos\left(\frac{\omega_Z(\tau-\tau')}{2}\right) e^{i\mathcal K^+({\bf r},T-\tau/2)+i\mathcal K^+({\bf r},T+\tau/2)-i\mathcal K^+({\bf r}',T-\tau'/2)-i\mathcal K^+({\bf r}',T+\tau'/2)-\frac{\tau-\tau'}{2\tau_B}}\\\times \int\limits_{{\bf x}(\tau')={\bf r}'}^{{\bf x}(\tau)={\bf r}}\mathcal D{\bf x}e^{-\int\limits_{\tau'}^\tau dt'
\left(\frac{(\dot{\bf x}(t'))^2}{2D}-\frac{i}{2}(\Phi^+({\bf x}(t'),T-t'/2)-\Phi^+({\bf x}(t'),T+t'/2))\right)},
\end{multline}
\end{widetext}
where $\tau_B=\frac{3c^2}{Da^2e^2B^2}$ is the well known expression for the electron decoherence time due to an external magnetic field.
Note that within this approximartion the symmetric Cooperon $\mathcal C_{s}$ turns out to be identical to $\mathcal C_{as}$. The difference
between these two objects (encountered, e.g. within the perturbation theory) shows up only in the subleading terms containing at least one of the quantum fields,
e.g., the $\Phi^-$-field.  The above representation of the Cooperon in terms of a path integral over diffusive electron trajectories proves to be very useful,
as it enables one to easily perform the Gaussian average over the fluctuating electromagnetic field.

\section{Andreev conductance}

Let us now employ the above formalism and evaluate the current $I$ flowing across the insulating barrier between the superconductor and the normal metal.
It reads
\begin{equation}
 I= \frac{e}{2}\int_{\rm I} d^{2}{\bf r}\langle\delta S_A/\delta {\mathcal K}^-({\bf r})\rangle_\Phi.
\label{Ac}
\end{equation}
Assuming that our system is biased by an external voltage $V$
and neglecting all terms containing $\Phi^-$ and $\mathcal K^-$, we find
\begin{multline}
I=\frac{\pi T}{2\nu e^3(R_Ia^2)^2}\int_{\rm I} d^{2}{\bf
r}d^{2}{\bf r}'\int d\tau \frac{\cos(\omega_Z\tau/2)e^{-\frac{\tau}{2\tau_B}}}{\sinh(\pi T\tau)}
\\\times {\rm Im}\langle\mathcal P({\bf r},{\bf r}';t;\tau) e^{ieV\tau}\rangle_\Phi,
\label{Ap}
\end{multline}
where
\begin{widetext}
\begin{multline}
\mathcal P({\bf r},{\bf r}',\tau;t)=
\frac{\theta(\tau) e^{i\mathcal K^+({\bf r},t-\tau)-i\mathcal K^+({\bf r},t)} }{2}\int\limits_{{\bf x}(0)={\bf r}'}^{{\bf x}(\tau)={\bf r}}\mathcal D{\bf x}e^{-\int\limits_{0}^\tau dt'
\left(\frac{(\dot{\bf x}(t'))^2}{2D}-\frac{i}{2}(\Phi^+({\bf x}(t'),t-(t'+\tau)/2)-\Phi^+({\bf x}(t'),t+(t'-\tau)/2))\right)}.
\end{multline}
\end{widetext}
Performing a straightforward Gaussian average over
the $\Phi^+$-fields, after additional averaging over diffusive trajectories in the case of quasi-1d $N$-metal wires (see Fig. 1)
for the differential Andreev conductance $G(V)=dI/dV$ we obtain
\begin{multline}
G=\frac{\pi T}{4\nu e^2 R_{\rm I}^2}\int\limits_0^\infty d\tau^2\frac{\mathcal D(0,0;\tau)\cos(eV\tau)\cos(\omega_Z\tau/2)}{\sinh(\pi T\tau)}\\\times e^{-f(0,0,\tau)-\frac{\tau}{2\tau_B}}
\label{finres}
\end{multline}
with $\mathcal D(0,0;\tau)=\vartheta_2(0,e^{-\tau/\tau_D})/(2La^2)$, where $\vartheta_2$ is the second Jacobi theta-function and $\tau_D=2L^2/(\pi^2D)$
is the Thouless time. In the limit $B \to 0$ the above expression for $G$ reduces to one derived in our previous work \cite{SZK12}.

The function $f$ describes dephasing of Cooper pairs. In the interesting for us low temperature limit $\pi T\tau\ll 1$ this function reduces to
\begin{multline}
f(0,0,\tau)\simeq\frac{8}{g_w}\ln\left(\frac{\tau}{\tau_{RC}}\right)+\frac{\tau}{\tau_{\varphi}}
+\sqrt{\frac{\pi\tau\tau_c}{4\tau_{\varphi}^2}}\ln\left(\frac{\tau_c}{\tau}\right),
\label{flim}
\end{multline}
where $\tau_{RC}=RC$ and $C$ is an effective junction capacitance. The first term in Eq. (\ref{flim}) is due to spatially uniform fluctuations
of the scalar potential \cite{HHK,ZGalakt}.
The remaining terms in this equation originate from
non-uniform in space fluctuations in the normal metal wire and determine the Cooper pair decoherence time $\tau_\varphi=2\pi\nu a^2\sqrt{2D\tau_c}$
as well as the Cooper pair decoherence length $L_\varphi =\sqrt{D\tau_\varphi}$, where $\tau_c \sim l/v_F$ sets a short time cutoff \cite{GZ1,GZ3,GZ4}
and also $\tau_\varphi \gg \tau_{RC}$.
It is important to observe that -- that up to an unimportant prefactor of order one  -- this dephasing time $\tau_\varphi$ coincides with that for single electrons
evaluated, e.g., for the WL problem \cite{GZ1,GZ3,GZ4}.

\section{Results and conclusions}

Turning to concrete results it is worthwhile to first recall the expressions for the Andreev conductance derived in the absence of the magnetic field \cite{SZK12}. In that case an important physical ingredient of the problem is a competition between the two fundamental length scales, the temperature length $L_T\sim \sqrt{D/T}$ and the decoherence length $L_\varphi$. The smallest of these two scales limits the proximity effect and influences both the linear conductance the ZBA peak. E.g. in the limit $L\gg L_\varphi$ and $T=0$ one finds \cite{SZK12}
\begin{equation}
G_{0}(V)\simeq\frac{1}{\sigma
R_{\rm I}^2a^2}\frac{L_\varphi}{\sqrt{2\pi}}\left(\frac{4\tau_{RC}}{\tau_\varphi}
\right)^{8/g_w}{\rm Re}\frac{
\Gamma\left(\frac12-\frac{8}{g_w}\right)}{ (1+ieV\tau_\varphi)^{1/2-8/g_w} }.
\label{B0}
\end{equation}
\begin{figure}[h]
\includegraphics[width=0.99\columnwidth]{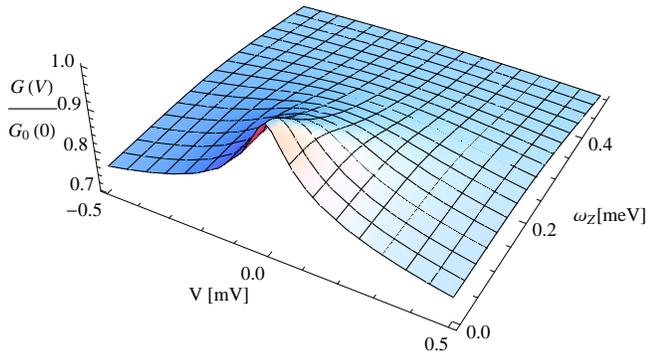}
\caption{$G(V)$ as a function
of external voltage $V$ and the magnetic field (in units of $\omega_Z$) for $a=10$ nm, $D=21$ cm$^2$/s. For these parameter values one finds
$1/\tau_\varphi \sim 0.6$ K and $L_\varphi \sim 0.2$ $\mu$m.} \label{fig4}
\end{figure}

Turning on the magnetic field $B$, from Eq. (\ref{finres}) we observe that its effect is twofold. Firstly, the magnetic field causes additional dephasing of Cooper pairs. This effect can be accounted for in Eq. (\ref{B0}) by the substitution
\begin{equation}
\tau_\varphi\to\frac{2\tau_B\tau_\varphi}{\tau_\varphi+2\tau_B}.
\end{equation}
Secondly, Zeeman splitting between states with opposite spins also influences
the subgap conductance. From Eq. (\ref{finres}) we obtain
\begin{equation}
G(V)=\frac12\left(G_0\left(V+\frac{\omega_Z}{2e}\right)+G_0\left(V-\frac{\omega_Z}{2e}\right)\right).
\end{equation}
This result implies that in the presence of the magnetic field the ZBA peak gets additionally smeared due to Zeeman splitting.  The behavior of the nonlinear conductance $G(V)$ in the presence of an external magnetic field and for $L > L_\varphi$ is displayed in Fig. 4. The dependence of the linear subgap conductance on the magnetic field is illustrated in Fig. 5.

\begin{figure}[h]
\includegraphics[width=0.99\columnwidth]{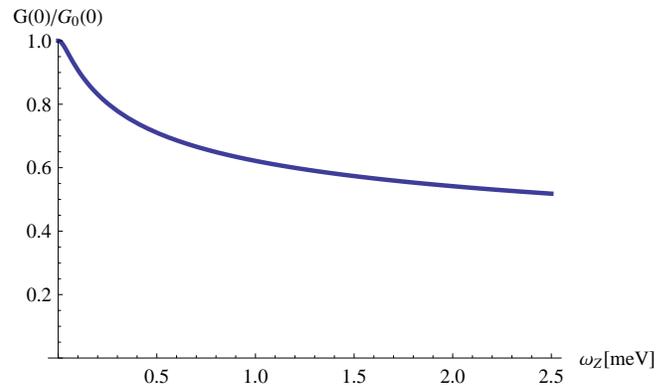}
\caption{The linear subgap conductance as a function of the magnetic field. The parameters are the same as in Fig. 4.} \label{fig5}
\end{figure}

Our results are qualitatively consistent with experimental findings \cite{Chand} which demonstrate an effective suppression of the low temperature magnetoconductance of SN structures with increasing magnetic field $B$. In this experiment the electron decoherence length  $L_\varphi$ was extracted from
independent weak localization measurements and was found to be temperature independent in the regime $L_\varphi \lesssim L_T$. The data \cite{Chand} also confirm that at low enough $T$ the subgap conductance of SN structures is determined by phase coherent electron paths with lengths restricted by the temperature independent value $L_\varphi$  rather than by the thermal length $L_T$ diverging in the low temperature limit.

In summary, we have demonstrated that even at $T \to 0$ electron-electron interactions yield effective dephasing of Cooper pairs penetrating from a superconductor into a diffusive normal metal. At low temperatures this phenomenon fundamentally affects the proximity effect in SN hybrids restricting the penetration length
of superconducting correlations into a normal metal to a $T$-independent value $L_\varphi$. In the presence of an external magnetic field the subgap conductance of SN structures gets reduced due to a combination of two effects -- additional temperature independent dephasing and Zeeman splitting between the states with opposite spins. Measurements of the subgap magnetoconductance in SN systems at low temperatures enable one to experimentally probe the fundamentally important parameter - dephasing length of Cooper pairs $L_\varphi$.

This work was partially supported by RFBR grant 12-02-00520-a.

\end{document}